\magnification=\magstep1
\baselineskip=13 pt plus 2pt minus 2pt
\vsize=8.9truein
\hsize=6.5truein
\footline={\ifnum\pageno=1 \hfil\else\centerline{\folio}\fi}

\font\smchapt=cmbx12
\font\abstrct=cmr8
\font\reglar=cmr10

\def\hi{\hangindent=15pt}
\def\vs{\vskip 6truept}
\def\vv{{\vs\vs}}
\def\veject{\vfill\eject}
\def\no{\noindent}
\def\h{\textstyle {1 \over 2}}
\def\p{\partial}
\def\bd{{\overline D}}
\def\bl{{\overline L}}
\def\wy{{\scriptscriptstyle \zeta}}
\def\bry{{{\overline{\wy}}}}

\def\bL{{\overline L}}
\def\sym{{\mathop{\otimes}\limits_{s}}\,}
\def\qcomma{\quad,\quad}
\def\qqcomma{\qquad,\qquad}

\def\by{{\zeta}}
\def\bby{{\overline{\zeta}}}
\def\E#1{Eq.~(#1)}

\smchapt
\centerline{An Iterative Approach to Twisting and Diverging, Type N,}
\centerline{Vacuum Einstein Equations:}
\centerline{~~A (Third-Order) Resolution of Stephani's `Paradox' }
\vv \reglar
\leftline{\qquad\qquad J. D. Finley, III\dag, J. F. Pleba\'nski\ddag${}^*$, 
 and Maciej Przanowski\S}\abstrct \vskip3pt
\leftline{\qquad\qquad~~\dag Department of Physics and Astronomy, 
University of New Mexico, Albuquerque,
N.M.  87131}
\leftline{\qquad\qquad~~\ddag Departamento de F\'isica, CINVESTAV del IPN, 
Apdo. Postal 14-740, 07000 M\'exico DF, M\'exico}
\leftline{\qquad\qquad~~\S Institute of Physics, Technical University, 
93-005 
\L \'od\'z, Poland}\vv
\centerline{\it Abstract}
\baselineskip=12pt {\narrower  \smallskip
In 1993, a proof was published, within this journal, that there are no 
regular 
solutions to the {\it linearized} version of the twisting, type-N, vacuum 
solutions of the Einstein field equations.  While this proof is certainly 
correct, we show that the conclusions drawn from that fact were 
unwarranted, 
namely that this irregularity caused such solutions not to be able to 
truly describe pure gravitational waves. In this article, we resolve the 
paradox---since such first-order solutions must always have singular lines 
in space for all sufficiently large values of $r$---by showing that if we 
perturbatively iterate the solution up to the third order in small 
quantities, there are acceptable  regular solutions.   That these solutions 
become flat before they become non-twisting tells us something 
interesting concerning the general behavior of solutions describing 
gravitational radiation from a bounded source.  \smallskip }
\baselineskip=20pt plus 2pt minus 2pt
\vv\reglar
\leftline{{\bf 1.~Introduction.~~}}\vv
The generic behavior of gravitational radiation from a bounded source is 
clearly an important physical problem.  Even reasonably far from that 
source, however, type-N solutions of the vacuum field equations must have 
both non-zero twist as well as reasonable asymptotic behavior in order to 
provide an exact description of that radiation.  Such solutions would 
provide small laboratories to better understand the complete nature of 
the singularities of type-N solutions, and could also be used to check 
numerical solutions that include gravitational radiation.  It is therefore 
reasonable that there is considerable interest in 
this problem.  In addition to non-zero values for the twist parameter, 
interesting solutions must also have appropriate asymptotic behavior; 
the only currently-known 
solution to the twisting problem, due to Hauser [1], does not have this 
asymptotic 
behavior.  As well the definitive relevance of homothetic vectors to the 
study of 
metrics of Petrov type N has been very well enunciated by McIntosh [2].  
It is the 
existence of two homothetic vectors ---more exactly, an $H_2$ of 
symmetries, a 
non-Abelian group of (local) homothetic vectors for the manifold---that 
allows 
the defining equations to be reduced to an ordinary differential equation.  
By construction, this equation must contain a constant parameter, the 
homothetic parameter, which McIntosh called $n$. Hauser's solution has 
the value of 5/2 for the McIntosh parameter, $n$.
\vs
Although several distinct formulations of this problem already exist, 
none have yet been able to produce new solutions.  We therefore believe 
that any exact solutions to this problem would be important in obtaining 
a better understanding of the general problem.  Clearly Hans Stephani felt 
much the same way when he published his proof  [3] that all, non-flat, 
{\it first-order} solutions of the twisting, type-N, vacuum Einstein 
field equations must always contain {\it singular lines}, i.e., 
places on the $\zeta,\bby$-sphere such that 
for sufficiently large values of $r$, and $u$, the analytic functions 
contained within the solution must a) be non-constant since otherwise 
the solution would be flat, and b) therefore must contain poles on that 
sphere which extend to infinite values of the affine parameter, $r$.  
Since it is true that such singular behavior is the rule for non-twisting 
Petrov type N solutions,  Stephani then conjectured that all {\bf pure}, 
Petrov type N solutions of the vacuum field 
equations would be insufficient to completely describe the propagation, 
in vacuum, of gravitational waves from a bounded source.  We resolve this 
apparent 
paradox, 
in the discussion below, by defining an algorithm that allows us to extend 
Stephani's discussion of the original Einstein field equations to an 
arbitrary 
order.  We then find that at third-order we can indeed show the existence 
of a 
solution that is non-flat (of Petrov type N), twisting, {\bf and} also 
regular 
for sufficiently large spheres.
\vs
\leftline{{\bf  2.~The description of the vacuum field equations:~~}}
\vs To most easily compare the equations to other forms, we give our 
presentation 
in the second variant of the usual null
tetrad formalism as originated by Debney, Kerr, and Schild in Section 4 of 
their paper [4]; however, for ease of comparison to the work of Stephani, 
we (mostly) use the same symbols as he does, which come originally from the 
work of Kramer, Stephani, MacCallum and  Herlt [5]. Therefore we write the 
metric, $\bf g$, in terms of a complex null tetrad, $e^\mu$ as follows:
$${\bf g} = g_{\mu\nu}e^\mu\sym e^\nu = 2e^1\sym e^2 + 2e^3\sym e^4\qqcomma 
\overline{e^1}
= e^2\,,\; e^3,\ e^4\  \hbox {~real,}\eqno(2.1)$$
\no and use the overbar for complex conjugation.  For
any vacuum, type-N space-time with
non-vanishing complex expansion, i.e., where $\, Z \equiv -\Gamma_{421}
\ne 0$, Debney, Kerr, and Schild 
showed the existence of local coordinates $\{\zeta,\bby\}$, complex, and  
$\{r,u\}$, real, with $r$ the affine parameter along the radiation 
trajectories.  
In terms of these coordinates, they showed that one can always write 
the null 
tetrad, $e^\mu$, so that 
$$\eqalign{e^1  = {1\over PZ}d\zeta\qcomma e^2 = & {1\over 
P\overline{Z}}d\bby  
\>,\quad e^3 = du + L\,d\zeta + \bL\, d\overline{\zeta}\quad,\cr
e^4 & = dr + W \,d\zeta +\overline{W}\,d\overline{\zeta}\  - \ H\,e^3
\quad,\cr}
\eqno(2.2)$$
\no where the metric functions are given by 
$$ \matrix{Z^{-1} = (r - i\Sigma)\,, & 2i\Sigma = P^2(\bd L - D\bl) \cr\cr
W = -{1\over \overline{Z}}L_{,u} + iD\Sigma\,, & D\equiv \p_\wy - L\,
\p_u \cr\cr
H = -r\p_u(\ln P) + \h K\,, & K = 2P^2\,\hbox{Re}[D(\bd \ln P - L_{,u})]
\cr}\eqno(2.3)$$ 
\no where the subscripts indicate partial differentiation.  Within this 
tetrad, 
setting $P\equiv V_{,u}$,  the
remaining Einstein vacuum field equations take the form
$$\eqalignno{\bd\bigg\{P^{-1} & \p_uDDV\bigg\} {}= 0\quad,&(2.4a)\cr
\noalign{\vskip-3truept}
\bd\,\bd D D V &{}= DD\bd\,\bd V\quad.&(2.4b)\cr}$$
Contrariwise, we insist that the solutions be non-flat, and have non-zero 
twist, which insists that both the following two quantities should not 
vanish:
$$\eqalign{ \overline{C^{(1)}}\  \propto\ &  \p_u\p_u\left\{P^{-1}\bd\bd 
V\right\} \ne 0\quad,\cr
2i\,\Sigma =  & P^2\left(\bd L - D\bL\right) \ne 0\;.\cr}\eqno(2.5)$$
\veject \leftline{{\bf  3.~The description of the perturbation 
equations:~~}}
\vs
To define the perturbation scheme, we first notice [5] that one may 
always take 
a gauge condition designed to maintain the function $P$ at its simplest 
value, 
namely 
$$ P = 1 + \h\zeta\bby\;.\eqno(3.1)$$
\no As well, if we now append the requirement that 
$L\equiv L(\zeta, \bby, u) = 0$, we get exactly the following form of 
the flat 
space-time metric:
$$ ds^2 = 2\left\{{r\over 1+\h\zeta\bby}\right\}^2d\zeta\sym d\bby +2\,
du\sym dr 
+ du\sym du\quad,\eqno(3.2)$$
\no where $\zeta$ and $\bby$ simply define the (usual) stereograhic 
coordinates 
on the complex sphere, while $r$ is the radial coordinate and $u$ is 
the retarded 
time coordinate.  Our perturbative scheme can then be developed by 
thinking of $L$ as 
being defined in terms of a repetitive sequence of approximations, 
involving higher 
and higher order approximations, and defining 
$$ \Phi\ \equiv \ -P^{-1}DDV\quad = \quad L_\wy + 
{\bby\over 1+\h\by\bby}L - 
L\p_u L \quad,\eqno(3.3)$$
\no in terms of them.  Therefore, \E{2.4a} now take the form 
$$\bd\p_u\Phi = 0 = \p_\bry\p_u\Phi - \bl\p_u^2\Phi \quad.\eqno(3.4)$$
\vs
It is now obvious that $L^{(0)} = 0$ implies that $\Phi^{(0)} = 0$, so that 
we may start with the lowest-order approximation, denoted in this form.  To 
find the next step---the first-order approximation---we suppose that both 
$L$ and $\Phi$ may be expanded into series, and take \E{3.3} as defining 
$\Phi$ for us in terms of $L$, and then take \E{3.4} as the 
constraint which 
defines $\Phi$ within the next order, so that we immediately  
have the constraint 
$$\p_\bry\p_u\Phi^{(1)} = 0\ \Longrightarrow \ \Phi^{(1)} = \alpha^{(1)}
(\by, u) + \beta^{(1)}(\by, \bby)\quad.\eqno(3.5)$$
\no Inserting this expression back into \E{3.3}, we obtain a
 defining relation for $L^{(1)}$, namely 
$$ \p_\wy L^{(1)} + {\bby\over 1+\h\by\bby} L^{(1)} = 
\Phi^{(1)} + L^{(0)}\p_u 
L^{(1)} + L^{(1)}\p_u L^{(0)} = \alpha^{(1)}(\by, u) + 
\beta^{(1)}(\by, \bby)
\quad.\eqno(3.6)$$
\vs 
The general solution of this equation, for $L^{(1)}$, is given by 
$$ L^{(1)} = (1+\h\by\bby)^{-2}\left\{f^{(1)}(\bby, u) + 
\int d\by\,(1+\h\by\bby)^2\,\left[\alpha^{(1)}(\by, u) + 
\beta^{(1)}(\by, \bby)\right]\right\}\quad.\eqno(3.7)$$
This allows us to write, now, the second-order steps in the form 
$$ \eqalign{\p_\bry\p_u\Phi^{(2)}  & = \bl^{(1)}\p^2_u\Phi^{(1)}\;,\cr
\p_\wy L^{(2)} + {\bby\over 1+\h\by\bby} L^{(2)}  & = \Phi^{(2)} + L^{(1)}
\p_u L^{(1)}\;.\cr}\eqno(3.8)$$
The general solution of these equations is then easily given in the form 
$$ \eqalign{\Phi^{(2)} = &  \int du\,\int d\bby\; \overline{L^{(1)}} 
\p_u^2\Phi^{(1)}\;,\cr
L^{(2)} = &\ (1+\h\by\bby)^{-2}\left\{ f^{(2)}(\bby,u) + 
\int d\by\,(1+\h\by\bby)^2\left[\Phi^{(2)} + L^{(1)}\p_u 
L^{(1)}\right]\right\}\;.\cr}\eqno(3.9)$$
\vs 
At the $n$-th level, the equations to be solved are then just 
$$\eqalign{\p_\bry\p_u\Phi^{(n)} =\  & \sum_{j=1}^{n-1}\,\
bl^{(n-j)}\p_u^2\Phi^{(j)}\;,\cr
\p_\wy L^{(n)} + {\bby\overwithdelims()1+\h\by\bby}\,L^{(n)} 
= &\ \Phi^{(n)}
 + \sum_{j=1}^{n-1} L^{(n-j)}\p_u L^{(j)}\;.\cr}\eqno(3.10)$$
The general ($n$-th order) solution of the equations for $\Phi$ 
and $L$ are 
then given by the series, up to the $n$-th order, of the $\Phi^{(j)}$ 
and the $L^{(j)}$, determined as 
$$ \eqalign{\Phi^{(n)} = \ & \sum_{j=1}^{n-1}\int du\;\int d\bby 
\;\bl^{(n-j)}\p_u^2\Phi^{(j)}\;,\cr
L^{(n)} =\ & (1+\h\by\bby)^{-2}\left\{f^{(n)}(\bby, u) + 
\int d\by\,(1+\h\by\bby)^2\left[\Phi^{(n)} + \sum_{j=1}^{n-1} 
L^{n-j}\p_u L^{(j)}\right]\right\}\;.\cr}\eqno(3.11)$$
\vs It now remains to consider \E{2.4b} which, in the current 
notation, can be written 
$$ \hbox{Im}(\bd\bd P\Phi) = 0\quad.\eqno(3.12)$$
Expanding out the terms in $\bd$, we obtain 
$$\eqalign{P\,\hbox{Im}\Big\{\p_\bry\p_\bry\Phi + {\by\over 1+\h\by\bby}\,
\p_\bry\Phi - \bl \p_u\p_\bry \Phi &  - \bl\left(\p_u\p_\bry\Phi - \bl 
\p_u^2\Phi\right) \cr 
& - \left(\p_\bry \bl + {\by\over 1+\h\by\bby} \bl - \bl\p_u\bl\right)
\p_u\Phi\Big\} = 0\;.}\eqno(3.13)$$
However, we can use \E{3.3} and \E{3.4} to re-write substantially the 
terms in the above equation, giving us 
$$ \hbox{Im}\left\{\p_{\bry}\p_\bry\Phi + {\by\over 1+\h\by\bby}\p_\bry
\Phi - \overline{\Phi}\p_u\Phi - \bl \p_\bry\p_u\Phi\right\} = 0\;.
\eqno(3.14)$$
Therefore, applying the same procedures as we have been using so far, 
for a constraint on the $n$-th iterative step, we find 
the linear, inhomogeneous, pde for $\Phi^{(n)}$, namely
$$ \hbox{Im}\left\{\p_\bry\p_\bry\Phi^{(n)} + {\by\over 1+\h\by\bby}\,
\p_\bry
 \Phi^{(n)}\right\} = \sum_{j=1}^{n-1}\hbox{Im}\left\{\overline
{\Phi^{(n-j)}}
\p_u\Phi^{(j)} + \overline{L^{(n-j)}}\p_\bry\p_u\Phi^{(j)}\right\}
\;.\eqno(3.15)$$
This equation surely does have solutions, and we may conclude that 
our iterative procedure is indeed complete.  Within the $n$-th order 
approximation, 
one has that
$$\Phi \approx \sum_{j=1}^n\,\Phi^{(n)} \;,\quad L \approx 
\sum_{j=1}^n\,L^{(n)}\quad.\eqno(3.16)$$
\vv\leftline{{\bf 4.~The first few approximate solutions~~}}\vs
Returning, first, to the linearized, or first-order approximation, we of 
course have that the solution is given by \E{3.5} and \E{3.7}, where 
$\alpha^{(1)} = \alpha^{(1)}(\by,u)$ and $f^{(1)} = f^{(1)}(\bby, u)$,
 while the function $\beta^{(1)} = \beta^{(1)}(\by,\bby)$ is allowed to 
be an solution of the equation 
$$ \hbox{Im}\left\{\p_\bry\p_\bry\beta^{(1)} + {\by\over 1+
\h\by\bby}\,\p_\bry \beta^{(1)}\right\} = 0\;.\eqno(4.1)$$
\no Defining, now, the function $E = E(\by,u)$ such that 
$$ \p_\wy^{\,3} E(\by,u) = \alpha^{(1)}(\by,u)\quad,\eqno(4.2)$$
we can quickly re-write $L^{(1)}$ in the form 
$$\eqalign{ L^{(1)} = & B(\by,\bby) + {C(\bby, u)\over (1+\h\by\bby)^2} + 
{\bby^2 E(\by, u)\over 2(1+\h\by\bby)^2} - {\bby \p_\wy E(\by, u)\over 1+
\h\by\bby} + \p^2_\wy E(\by, u) \quad,\cr
&\qquad \hbox{where}~B(\by,\bby) \equiv {1\overwithdelims() 1+\h\by\bby}^2
\,\int d\by \;(1+\h\by\bby)^2\,\beta^{(1)}(\by,\bby)\,,\cr}\eqno(4.3)$$
and we have denoted our $f^{(1)}(\bby, u)$ by Stephani's symbol 
$C(\bby, u)$, 
while we have used $E(\by, u)$ to denote his $D(\by,u)$.  With these 
identifications, this is exactly the solution given by Stephani in [3].   
Insisting that the solution be non-flat, we infer that it is necessary 
for the existence of a non-trivial, linearized Petrov type N solution 
that we must have 
$$ \p^2_u\p_\bry{}^3 \overline{E} \ne 0\quad.\eqno(4.4)$$
However, as has already been pointed out by Stephani [3], this can happen 
if and only if the field is singular at some point on the 
$\by,\bby$-sphere.  
Every non-singular field defines a flat spacetime, only, 
at this level of approximation.  
\vs To see this in some more detail, we note, still following Stephani, 
that the invariants available to us are 
$$\eqalign{{\cal J}_1 \approx {\cal J}_1^{(1)} = &  {P^2\over 2ir}\left
\{\p_\bry L^{(1)} - \p_\wy \bl^{(1)}\right\}\;,\cr
{\cal J}_2 \approx {\cal J}_2^{(0)} + {\cal J}_2^{(1)}  & = {1\over r^2}
\left\{ 1 - P^2(\p_\wy\p_u\bl^{(1)} + \p_\bry\p_u L^{(1)})\right\}\;.\cr}
\eqno(4.5)$$
Differentiating the first of these with respect to $u$, and comparing the 
result with the second, one easily concludes that if both of these are 
regular on the $\by,\bby$-sphere, then it must also be true that $P^2\p_
\bry\p_u L^{(1)}$ must be regular there.  Employing our equation, \E{4.3}, 
for $L^{(1)}$, we easily calculate that this quantity is given by 
$$ P^2\p_\bry\p_u L^{(1)} = - {\by\over 1+\h\by\bby}\p_u C(\bby, u) + 
\p_\bry \p_u C(\bby, u) + {\bby\over 1+\h\by\bby}\p_u E(\by,u) 
- \p_\wy\p_u E(\by,u)\;.\eqno(4.6)$$
\vs
We easily infer that this quantity is, then, regular on the sphere, if and 
only if the functions $C$ and $D$ are of the form 
$$ \eqalign{C(\bby,u) =  & \bby\,r(u) + \sigma(u) + \rho(\bby)\,,\cr
E(\by,u) = & \by\,\delta(u) + \omega(u) + h(\by)\,.\cr}\eqno(4.7)$$
However, the form given in \E{4.7} for $E(\by,u)$ would obviously lead to 
the space-time being flat, rather than, non-trivially, of Petrov type N.  
This result, given by Stephani [3], seems to lead to the following 
conclusion, 
which we refer to as Stephani's `paradox,' namely [3]\par
{\narrower So either twisting type N fields do {\bf not} describe a 
radiation 
field outside a bounded source, or due to a mechanism not yet recognized, 
they cannot be linearized, or the na\"ive interpretation of the coordinates 
(starting from their Newtonian limit) is wrong.\smallskip}
\vs \no
The most catastrophic of these options is the first possibility, 
i.e., that  
``twisting, type N fields do {\bf not} describe a radiation field 
outside a bounded source.''  In the next section, we find a solution 
for the twisting, 
type N field, at the third step of iteration.  Our solution is in fact 
{\bf regular} on the $\by,\bby$-sphere.  This allows us to resolve the 
catastrophic 
argument cited above, by saying that it is indeed not true.
\vv\leftline{{\bf 5.~Example of a regular solution at the third level of 
approximation.}}\vs
Referring back to \E{4.3} for $L^{(1)}$, we choose it to have the 
very simple form
$$ L^{(1)} = {a^{(1)}\over \by}\;,
\quad \Rightarrow \ \Phi^{(1)} = {a^{(1)}\over \by}\left\{{\bby\over 
1+\h\by\bby} - {1\over \by}\right\}\;,
\qquad a^{(1)}~\hbox{ a complex constant.}\eqno(5.1)$$
It is straightforward to check that this solution does indeed satisfy 
\E{3.15} for $n = 1$.  Now, we may choose---see \E{3.9}---the second-order 
quantities to be given by 
$$ L^{(2)} = {a^{(2)}\over \by} + {\bby f^{(2)}(u)\over 
(1+\h\by\bby)^2}\;,\quad 
\Rightarrow\quad \Phi^{(2)} = {a^{(2)}\over \by}\left\{{\bby
\over 1+\h\by\bby} - {1\over \by}\right\}\;.\eqno(5.2)$$
Once again, it is evident that the constraint \E{3.15} is satisfied for 
$n=2$.  Now, using the $n$-th order equation for $n=3$, we find that 
$$ L^{(3)} ={1 \over (1+\h\by\bby)^2}\left\{\bby \,f^{(3)}(u) + \int d
\by\left[ (1+\h\by\bby)^2\Phi^{(3)} + {\bby\over \by} a^{(1)}\,{d^2f^
{(2)}(u)\over du^2}\right]\right\}\;.\eqno(5.3)$$
We may then set $a^{(3)}$ as yet another complex constant, and put 
$$ \Phi^{(3)} = - {a^{(1)}\over \by^2}{d^2f^{(2)}(u)\over du^2} + 
{a^{(3)}\over \by}\left\{{\bby \over 1+\h\by\bby} - {1\over \by}\right\}
\,.\eqno(5.4)$$
\vs
Once again, we may substitute this value for $\Phi^{(3)}$ into \E{5.3} 
for $L^{(3)}$, which gives us its form:
$$ L^{(3)} = {\bby\,f^{(3)}(u)\over (1+\h\by\bby)^2} + {a^{(1)}\over \by}
{d^2f^{(2)}(u)\over du^2}\, {1-\h\by\bby\over 1+\h\by\bby} 
+ {a^{(3)}\over \by}\;.\eqno(5.5)$$
Inserting, as before, these values into \E{3.15}, we find that they 
satisfy that equation for $n=3$.  
\vs
We must now determine the values for the twist and the curvature that 
go along with this, third-order approximation for the complete solution.  
We have, immediately, that the twist, $\Sigma$, is given by 
$$\eqalign{\Sigma^{(0)} \ & = \ {}~0 \,,\cr
\Sigma^{(1)}\ &  =\ {1\over 2i}P^2\left(\p_\bry L^{(1)} - \p_\wy 
\bl^{(1)}\right) = 0\,,\cr
\Sigma^{(2)} \ & = \ {1\over 2i}P^2\left(\p_\bry L^{(2)} - \p_\wy 
\bl^{(2)}\right) =  {1-\h\by\bby\over 1+\h\by\bby}\,\hbox{Im}
(f^{(2)}(u))\;,\cr
\Sigma^{(3)} \ & =\ {1\over 2i}P^2\left(\p_\bry L^{(3)} - 
\overline{L^{(1)}}\p_u
L^{(2)} - \p_\wy \bl^{(3)} + L^{(1)}\p_u\overline{L^{(2)}}
\right) \cr
& \qquad =  ~{1-\h\by\bby\over 1+\h\by\bby}\,\hbox{Im}(f^{(3)}(u)) - 2
[\hbox{Re} (a^{(1)})]\hbox{Im}\left({df^{(2)}(u)\over du}\right)\;,\cr}
\eqno(5.6)$$
along with the iterated values for the contributions 
to the (2-dimensional) curvature, $K$:
$$\eqalign{K^{(0)} = 2P^2\hbox{Re}(D\bd\ln P) = & {}~1\,,\cr
K^{(1)} = -2P^2\hbox{Re}(\p_\wy\p_u\bl^{(1)}) =  &{} ~0\,,\cr
K^{(2)} = -2P^2\hbox{Re}(\p_\wy\p_u\bl^{(2)}) =  & {}~-2{1-\h
\by\bby\over 1+\h\by\bby}\,\hbox{Re}\left({df^{(2)}(u)\over du}
\right)\;,\cr
K^{(3)} = -2P^2\hbox{Re}(\p_\wy\p_u\bl^{(3)} - L^{(1)}\p_u^2\bl
^{(2)}) = & {} 
~-2{1-\h\by\bby\over 1+\h\by\bby}\,\hbox{Re}\left({df^{(3)}(u)
\over du}
\right)\cr
&{}~~ + 4\>[\hbox{Re} (a^{(1)})]\hbox{Re}\left({d^2f^{(2)}(u)\over 
du^2}\right)\;,\cr}\eqno(5.7)$$ 
\no so that the total, 4-dimensional curvature, $\overline{C^{(1)}}\equiv 
2\Psi_4$, is given iteratively by 
$$ \eqalign{\Psi_4^{(0)} = & ~0\,,\quad \Psi_4^{(1)} = 
0\,,\quad \Psi_4^{(2)} = 0\,,\cr
\Psi_4^{(3)} = &  {\overline{a^{(1)}}\over r}{1+\h\by\bby
\overwithdelims() \bby}^2{d^3f^{(2)}(u)\over du^3}\,.\cr}\eqno(5.8)$$
\vs
Bringing together the sum of all these quantities, to write the complete 
solution valid up through the third iteration step, we have
$$\eqalign{\Phi \ \approx\  & {a_1\over \by}\left({\bby\over 1+\h\by\bby}
 - {1\over \by}\right) - {a_1\over \by^2}{df_2(u)\over du}\;,\cr
L\ \approx\ & {a_1\over \by} + {\bby\over (1+\h\by\bby)^2}f_2(u) + 
{1-\h\by\bby\over 1+\h\by\bby}{a_1\over \by}{df_2(u)\over du}\;,\cr
\Sigma \ \approx \ & {1-\h\by\bby\over 1+ \h\by\bby}\hbox{Im}[f_2(u)] 
- 2\hbox{Re}(a_1)\hbox{Im}\left({df_2(u)\over du}\right);,\cr
K\ \approx \ & 1 - 2{1-\h\by\bby\over 1+\h\by\bby}\hbox{Re}\left({df_2(u)
\over du}\right) + 4[\hbox{Re} (a_1)]\hbox{Re}\left({d^2f_2(u)\over 
du^2}\right)\;,\cr}\eqno(5.9)$$
\no along with the curvature itself, 
$$\Psi_4 \ \approx \ {\overline{a_1}\over r}{1+\h\by\bby
\overwithdelims() \bby}^2{d^3\overline{f_2}(u)\over du^3}\;,
\eqno(5.10)$$
\no where we have defined 
$$ a_1 \equiv a^{(1)} + a^{(2)} + a^{(3)}\;,\qquad f_2(u) 
\equiv f^{(2)}(u) + f^{(3)}(u)\;.\eqno(5.11)$$
\vs In order to actually determine the metric itself, we must, 
lastly, determine the function, $W$, which, to this level of 
iteration, is given by 
$$ W \ \approx \ -{\bby\over (1+\h\by\bby)^2}\left(r{df_2(u)\over du} + 
i\hbox{Im} f_2(u)\right) -{a_1\over\by}{1-\h\by\bby\over 1+\h\by\bby}
\left(r{ d^2f_2(u)\over du^2} + i\hbox{Im}{df_2(u)\over du}\right)\;.
\eqno(5.12)$$
Inserting all this into the equation for the metric, $\bf g$, itself, 
we see that the metric does indeed appear to be everywhere regular on 
the $\by,\bby$-sphere.  To completely show this, one needs only to 
change the coordinates in a neighborhood of the north pole, according 
to the usual rule, $\by' \equiv 1/\by$, and $\bby'\equiv 1/\bby$, 
which causes no trouble at all.  
\vs 
Gathering all our results together, we conclude that indeed one can find 
a {\bf regular, twisting, non-flat, Petrov type N, vacuum} metric that is 
regular on the $\by,\bby$-sphere in the third order of approximation.  
Consequently it seems that the twisting, type N fields {\bf can} describe 
a radiation field outside a bounded source.  Of course it is quite 
interesting to determine what happens in the next iteration steps.  
We intend to consider this question soon.
\vs
\no{{\it Acknowledgments:}}~~Thanks need also to be extended to Mark 
Hickman who assisted with the final form of the presentation of this 
work.\vs
\centerline{REFERENCES}
\parindent=0pt
\baselineskip=14pt
\def\hi{\hangindent=20truept}
\def\bn#1{{\bf #1}}

\frenchspacing
\vs

\hi ${}^*$ On leave of absence from the University of 
Warsaw, Warsaw, Poland.

\hi[1]  I. Hauser, Phys. Rev. Lett \bn{33} (1974) 1112, 
J. Math. Phys.\bn{19} (1978) 661.

\hi [2]  McIntosh, C.B.G., Classical \& Quantum Gravity \bn 2 (1985) 87-97.

\hi[3] H. Stephani, Classical \& Quantum Gravity \bn {10} (1993) 2187-2190.

\hi [4]  G. C. Debney, R. P. Kerr and A. Schild, J. Math. Phys. \bn{10} 
(1969) 1842-1854.

\hi[5]  D. Kramer, H. Stephani, M. MacCallum and E. Herlt, {\it Exact
solutions of Einstein's field equations}, (VEB Deutscher Verlag der
Wissenschaften, Berlin; Cambridge Univ. Press, Cambridge, 1980).

\vfill
\eject
\end
\bye